\documentclass[aps,prl,showpacs,preprintnumbers,twocolumn,nkgroupedaddress,amsfonts]{revtex4}
\usepackage{graphicx}
\usepackage{dcolumn}
\usepackage{bm}
\usepackage{epsfig}
\usepackage{amssymb,amsmath}
\suppressfloats[!]

\def \ptgg  {$p_T^{\gamma\gamma}$}
\def \mgg  {$M_{\gamma\gamma}$}
\def \dphigg {$\Delta\phi_{\gamma\gamma}$}
\def \cosgg {$|\cos \theta^{*}|$}

\def \Dptgg  {$d\sigma/d$\ptgg}
\def \Dmgg  {$d\sigma/d$\mgg}
\def \Ddphigg {$d\sigma/d$\dphigg}
\def \Dcosgg {$d\sigma/d$\cosgg}

\def \DTptgg  {$d^{2}\sigma/d$\mgg$d$\ptgg}
\def \DTdphigg {$d^{2}\sigma/d$\mgg$d$\dphigg}
\def \DTcosgg {$d^{2}\sigma/d$\mgg$d$\cosgg}

\begin{document}

\hspace{5.2in}\mbox{FERMILAB-PUB-10-039-E}

\title{\boldmath Measurement of direct photon pair production cross sections
in $p\bar{p}$ collisions at $\sqrt{s}=1.96$~TeV}

% LIST_OF_AUTHORS_R2.TEX                  2/4/10            
%
\author{V.M.~Abazov$^{36}$}
\author{B.~Abbott$^{74}$}
\author{M.~Abolins$^{63}$}
\author{B.S.~Acharya$^{29}$}
\author{M.~Adams$^{49}$}
\author{T.~Adams$^{47}$}
\author{E.~Aguilo$^{6}$}
\author{G.D.~Alexeev$^{36}$}
\author{G.~Alkhazov$^{40}$}
\author{A.~Alton$^{62,a}$}
% ...

\affiliation{\vspace{0.1 in}(The D\O\ Collaboration)\vspace{0.1 in}}
% ...
\affiliation{$^{6}$Simon Fraser University, Burnaby, British Columbia, Canada;
                and York University, Toronto, Ontario, Canada}
% ...
\affiliation{$^{29}$Tata Institute of Fundamental Research, Mumbai, India}
% ...
\affiliation{$^{36}$Joint Institute for Nuclear Research, Dubna, Russia}
% ...
\affiliation{$^{40}$Petersburg Nuclear Physics Institute,
                St. Petersburg, Russia}
% ...
\affiliation{$^{47}$Florida State University, Tallahassee, Florida 32306, USA}
% ...
\affiliation{$^{49}$University of Illinois at Chicago,
                Chicago, Illinois 60607, USA}
% ...
\affiliation{$^{62}$University of Michigan, Ann Arbor, Michigan 48109, USA}
\affiliation{$^{63}$Michigan State University,
                East Lansing, Michigan 48824, USA}
% ...
\affiliation{$^{74}$University of Oklahoma, Norman, Oklahoma 73019, USA}
% ...

\date{May 12, 2010}

\begin{abstract}\noindent
We present a measurement of direct photon pair production
cross sections using 4.2 fb$^{-1}$ of data collected
with the D0 detector at the Fermilab Tevatron $p\bar{p}$
Collider.
We measure single differential cross sections as a function of the diphoton
mass, the transverse momentum of the diphoton system, the azimuthal angle
between the photons, and the polar scattering angle of the photons.
In addition, we measure double differential cross sections considering
the last three kinematic variables in three diphoton mass bins.
The results are compared with different perturbative QCD predictions and event generators.
\end{abstract}

\maketitle

At a hadron collider, the direct photon pair (DPP) production with large diphoton invariant mass (\mgg)
cons\-ti\-tu\-tes a large and irreducible background to searches for the Higgs boson
decaying into a pair of photons, for both the Fermilab Tevatron \cite{hgg_prl} and the CERN LHC
experiments \cite{lhc}. DPP production is also a significant background in searches
for new phenomena, such as new heavy resonances~\cite{bsm1},
extra spatial dimensions~\cite{ED}, or cascade decays of heavy new particles~\cite{Susy}.
Thus, precise measurements of the diphoton differential production cross sections
for va\-rious kinematic variables and their
theoretical understanding are extremely important for future Higgs and new phenomena searches.

In addition, DPP production is interesting in its own right, and is used to check the validity
of the predictions of perturbative quantum chromodynamics (pQCD)
and soft-gluon resummation methods implemented in theoretical calculations.
Measurements involving the diphoton final state have been previously carried out
at fixed-target \cite{WA70,E706} and collider \cite{UA1,UA2,CDF} experiments.
However, the large integrated luminosity accumulated by the D0 experiment
in $p\bar{p}$ collisions at $\sqrt{s}=1.96$ TeV at the Fermilab Tevatron Collider
allows us to perform precise measurements of several observables in kinematic
regions previously unexplored, as well as, for the first time, the measurement
of double differential cross sections for this process.

\begin{figure}[htbp]
\centering
\hspace*{-9mm} \includegraphics[scale=0.41]{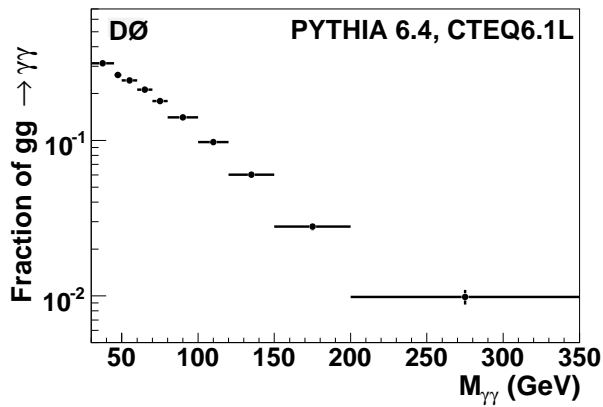}
~\\[-4mm]
\caption{The fraction of events produced via $gg \to \gamma\gamma$ scattering relative to total
diphoton production as a function of \mgg, as predicted by the {\sc pythia} event generator using the CTEQ6.1L PDF set.
Photons are required to have transverse momentum $p_T > 21(20)$ GeV
for the highest (next-to-highest) $p_T$ photon and pseudorapidity
$|\eta| < 0.9$ \cite{d0_coordinate}.}
\label{fig:ggF}
\end{figure}

The DPP events produced in $p\bar{p}\rightarrow \gamma \gamma+X$ are expected to
be dominantly produced via $q\bar{q}$ scattering ($q\bar{q} \rightarrow \gamma\gamma$) and
gluon-gluon fusion ($gg \rightarrow \gamma \gamma$) through a quark-loop diagram.
In spite of the suppression factor of $\alpha_s^2$ for  $gg\rightarrow \gamma\gamma$
as compared to $q\bar{q}\rightarrow \gamma\gamma$, the former
still gives a significant
contribution in kinematic regions where the $gg$ parton luminosity is high, especially at low \mgg.
Figure~\ref{fig:ggF} shows the expected contribution to the total DPP rate from $gg\rightarrow \gamma\gamma$,
as predicted by the {\sc pythia} \cite{pythia-ref} Monte Carlo (MC) event generator
with the CTEQ6.1L parton distribution function (PDF) set \cite{CTEQ}.
In addition, direct photons may result from single or double fragmentation processes of the partons
produced in the hard scattering \cite{resbos-ref,diphox-ref}. However,
a strict photon isolation requirement significantly reduces the rate for these processes.

In this Letter, we present measurements of the DPP production
cross sections using data collected by the D0 experiment from August 2006 to June 2009.
The cross sections are measured differentially
as a function of \mgg, the diphoton transverse momentum (\ptgg),
the azimuthal angle between the photons (\dphigg), and
the cosine of the polar scattering angle of the photon in the frame with
no net transverse momentum of the diphoton system
(defined as $\cos \theta^{*} = \tanh[(\eta_{1}-\eta_{2})/2]$, where $\eta_{1(2)}$
is the pseudorapidity of the highest (next-to-highest) $p_T$ photon).
These kinematic variables probe different aspects
of the DPP production mechanism. For instance,
the shapes of the \ptgg ~and \dphigg ~distributions are mostly affected by initial state gluon radiation
and fragmentation effects. In addition, the \mgg ~spectrum is particularly sensitive to potential contributions from new phenomena.
The $\cos \theta^{*}$ distribution probes PDF effects and the angular momentum of the final
state, which should be different for QCD-mediated production as compared, for example, to the
decay of a spin-0 Higgs boson~\cite{resbos-ref}.
The measured cross sections are compared to theoretical predictions from
{\sc resbos} \cite{resbos-ref}, {\sc diphox} \cite{diphox-ref}, and {\sc pythia} \cite{pythia-ref}.
Both {\sc resbos} and {\sc diphox} provide next-to-leading order (NLO) predictions in pQCD, however
the $gg \rightarrow \gamma \gamma$ contribution is considered only at leading order (LO) in {\sc diphox}.
{\sc pythia} is a parton shower MC event generator that includes the above processes at LO.
In {\sc diphox}, the explicit parton-to-photon fragmentation functions are included at NLO, while
in {\sc resbos} a function approximating rate from the NLO fragmentation diagrams is introduced.
Also, only in {\sc resbos}, the effects of soft and collinear
initial state gluon emissions are resummed to all orders. This is particularly important for the
description of the \ptgg~(\dphigg) distribution, which is a $\delta$-function at LO and
diverges at NLO as $p_T^{\gamma\gamma}\to 0$ ($\Delta \phi_{\gamma\gamma} \to \pi$).

The D0 detector is a general purpose detector
discussed in detail elsewhere~\cite{d0det}.
The subdetectors most relevant to this analysis are the central tracking
system, composed of a silicon microstrip tracker (SMT) and a central fiber
tracker (CFT) embedded in a 2~T solenoidal magnetic field, the central
preshower detector (CPS), and the calorimeter.
The CPS is located immediately before the inner layer of the calorimeter
and is formed of approximately one radiation length of lead absorber followed by three
layers of scintillating strips. The calorimeter consists of a central section with
coverage in pseudorapidity of $|\eta_{\rm det}|<1.1$~\cite{d0_coordinate},
and two end calorimeters covering up to $|\eta_{\rm det}| \approx 4.2$.
The electromagnetic (EM) section of the
calorimeter is segmented longitudinally into four layers (EM$i$, $i=1,4$), with transverse
segmentation into cells of size $\Delta\eta_{\rm det}\times\Delta\phi_{\rm det} = 0.1\times 0.1$~\cite{d0_coordinate},
except EM3 (near the EM shower maximum), where it is $0.05\times 0.05$.
The calorimeter is well-suited for a precise measurement of the energy
and direction of electrons and photons,
providing an energy resolution of about $3.6\%$ at an energy of $50$~GeV
and an angular resolution of about $0.01$ radians.
The energy response of the calorimeter to photons is calibrated using
electrons from $Z$ boson decays. Since electrons and photons shower
differently in matter, additional corrections as a function of $\eta$ are
derived using a detailed {\sc geant}-based \cite{GEANT} simulation
of the D0 detector response. These corrections are largest [($2.0 - 2.5$)\%]
at low photon energies ($\approx 20$ GeV).
The data used in this analysis were collected using a combination of triggers
requiring at least two clusters of energy in the EM calorimeter with
loose shower shape requirements and varying $p_T$ thresholds between 15 GeV
and 25 GeV, and correspond to an integrated luminosity of $4.2 \pm 0.3$~fb$^{-1}$~\cite{d0lumi}.

Events are selected by requiring two photon candidates with
transverse momentum $p_T>21$ ($20$)~GeV for the highest (next-to-highest) $p_T$
photon candidate and pseudorapidity $|\eta|<0.9$, for which the trigger
requirements are $>\!96\%$ efficient.
The minimum $p_T$ requirements for the two photon candidates are chosen to be different
following theoretical discussions \cite{resbos-ref,diphox-ref} and
a previous measurement \cite{CDF}.
The photon $p_T$ is computed with respect
to the reconstructed event primary vertex (PV) with the highest number of
associated tracks.
The PV is required to be within 60~cm
of the center of the detector along the beam axis. The PV has a
reconstruction efficiency  of about 98\% and has about 65\% probability of
being the correct vertex corresponding to the hard $p\bar{p}\to \gamma\gamma + X$ production.

Photon candidates are formed from clusters of calorimeter cells
within a cone of radius ${\cal R}=\sqrt{(\Delta\eta)^2+(\Delta\phi)^2}=0.4$ around a seed tower
\cite{d0det}.
The final cluster energy is then recalculated from the inner core with ${\cal R}=0.2$.
The photon candidates are selected by requiring:
(i) $\geq 97\%$ of the cluster energy be deposited in the
EM calorimeter layers; (ii) the calorimeter isolation
$\mathcal{I}=[E_{\text{tot}}(0.4)-E_{\text{EM}}(0.2)]/E_{\text{EM}}(0.2)<0.10$,
where $E_{\text{tot}}({\cal R})$ [$E_{\text{EM}}({\cal R})$] is the total [EM only]
energy in a cone of radius ${\cal R}$; (iii) the $p_T$ scalar sum of
all tracks originating from the PV in an annulus of $0.05<{\cal R}<0.4$ around the EM cluster
be $<1.5$~GeV; and
(iv) the energy-weighted EM shower width
be consistent with that expected for an electromagnetic shower.
To suppress electrons misidentified as photons,
the EM clusters are required to not be spatially matched to significant tracker activity,
either a reconstructed track or a density of hits in the SMT and CFT consistent
with that of an electron~\cite{HOR}. In the following, this requirement will be referred to as the ``track-match veto''.

\begin{figure}[htbp]
\centering
\hspace*{-9mm} \includegraphics[scale=0.41]{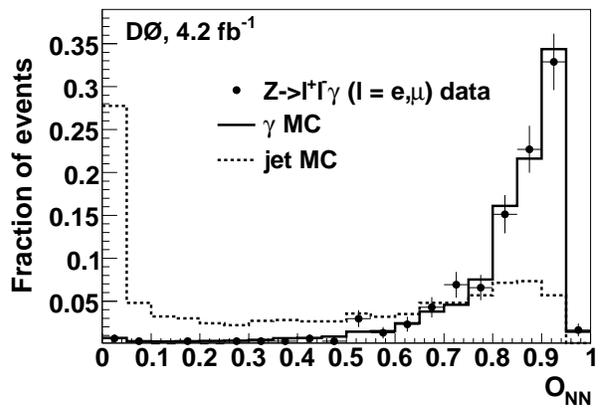}
~\\[-4mm]
\caption{Comparison of the normalized $O_{\rm NN}$ spectra for photons from DPP MC and
$Z\to\ell^+\ell^-\gamma$ data and for misidentified jets from dijet MC.}
\label{fig:NN_out}
\end{figure}

\begin{figure}[htbp]
\centering
\hspace*{-9mm} \includegraphics[scale=0.40]{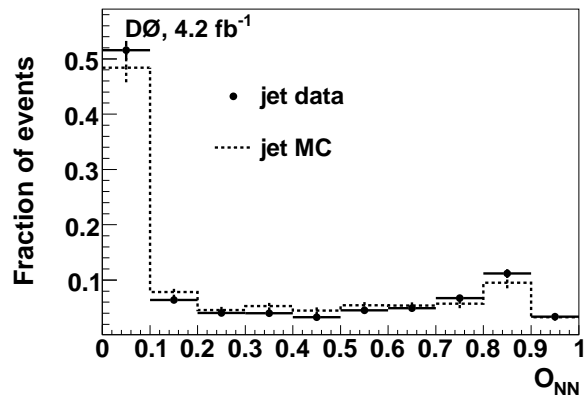}
~\\[-4mm]
\caption{Comparison of the normalized $O_{\rm NN}$ spectra for jets misidentified as photons
in data and in dijet MC.}
\label{fig:NN_out_jets}
\end{figure}

To further suppress jets misidentified as photons, an artificial neural network (NN) discriminant which
exploits differences in tracker activity and energy deposits in the calorimeter and in the CPS
between photons and jets is defined~\cite{hgg_prl}.
The NN is trained using $\gamma$ and jet {\sc pythia} MC samples. The shapes
of the NN output ($O_{\rm NN}$), normalized to unit area and obtained
after applying all data selection criteria, are shown in Figure \ref{fig:NN_out},
exhibiting a significant discrimination between photons and jets.
Photon candidates satisfy the requirement $O_{\rm NN} > 0.3$, which is $\approx\!98\%$ efficient
for photons and rejects $\approx\!40\%$ of the jets misidentified as
photons.
The $O_{\rm NN}$ shape is validated in data.
For photons a data sample consisting of photons radiated
from charged leptons in Z boson decays ($Z\to\ell^+\ell^-\gamma$, $\ell=e, \mu$)~\cite{Zg}
is used.
The MC modeling of the $O_{\rm NN}$ shape for jets is validated in a sample
of photon candidates selected by inverting the photon isolation (${\cal I} > 0.07$),
a requirement that significantly enriches the sample in jets.
The data and MC $O_{\rm NN}$ shapes are compared
in Figures \ref{fig:NN_out} and \ref{fig:NN_out_jets} and found to be in good agreement.

Finally, the two photon candidates are required to be spatially separated from
each other by a distance in $\eta-\phi$ space ${\Delta\cal{R}} > 0.4$ and to
satisfy  $M_{\gamma\gamma}>p_T^{\gamma\gamma}$.
The latter requirement is satisfied by the majority ($\approx\!92\%$) of DPP events and, together
with the photon isolation requirements, allows  significant suppression of the
contribution from the fragmentation diagrams, thus restricting the
data-to-theory comparison to the region where the theoretical calculations
should have smaller uncertainties \cite{resbos-ref}.

After imposing all requirements, 10938 events with diphoton candidates are selected in data. This sample includes
instrumental background contributions from $\gamma$+jet and dijet production, where a jet
is misidentified as a single photon as a result of fluctuations in the parton fragmentation
into a well-isolated neutral meson ($\pi^0$ or $\eta$) decaying into a final state with two or more photons.
An additional smaller background contribution results from $Z$-boson/Drell-Yan production events
$Z/\gamma^* \to e^+e^-$ (ZDY) in which both electrons are misidentified as photons.

The contribution from ZDY events is estimated using the MC simulation with {\sc pythia},
normalized to the NNLO cross section~\cite{ZDY_NNLO}. The selection efficiencies
determined from the MC simulation are corrected to those measured in
the data. On average, each electron has a 2\% probability of satisfying the photon selection
criteria, mainly due to the inefficiency of the track-match veto requirements.
The total ZDY contribution is estimated to be $161 \pm 20$ events.
Backgrounds due to $\gamma+$jet and dijet events are estimated from
data by using a $4\times4$ matrix background estimation method~\cite{hgg_prl}.
After applying all of the selection criteria described above,
a tighter $O_{\rm NN}$ requirement ($O_{\rm NN}>0.6$)
is used to  classify the data events into four categories, depending on whether
both photon candidates, only the highest $p_T$ one, only the next-to-highest $p_T$ one, or neither
of the two photon candidates pass ($p$) or fail ($f$) this requirement.
The corresponding number of events (after subtraction of the estimated ZDY contribution)
compose a 4-component vector ($N_{pp}, N_{pf}, N_{fp}, N_{ff}$).
The difference in relative efficiencies of the $O_{\rm NN}>0.6$ requirement between
photons and jets allows estimation of the sample composition by solving a linear
system of equations:
$(N_{pp}, N_{pf}, N_{fp}, N_{ff})^T = {\cal E}\times (N_{\gamma\gamma}, N_{\gamma j}, N_{j\gamma}, N_{jj})^T$,
where $N_{\gamma\gamma}$ ($N_{jj}$) is the number of DPP (dijet) events and
$N_{\gamma j}$ ($N_{j\gamma}$) is the number of $\gamma$+jet events with the
(next-to-)highest $p_T$ photon candidate being a photon. The $4\times 4$ matrix ${\cal E}$ contains
the photon $\varepsilon_{\gamma}$ and jet $\varepsilon_{\rm jet}$ efficiencies, estimated using
photon and jet MC samples and validated in data.
The efficiencies are parameterized as a function of the photon candidate $\eta$ and vary within ($90 - 95$)\% for
$\varepsilon_{\gamma}$  and within ($66 - 70$)\% for $\varepsilon_{\rm jet}$.
The systematic uncertainty on $\varepsilon_{\gamma}$ is estimated to be 1.5\% from
a comparison of the efficiency as a function of $\eta$ between data and MC using samples of
electrons from $Z$ boson decays and photons from radiative $Z$ boson decays. In order to
estimate the systematic uncertainty on $\varepsilon_{\rm jet}$, two independent control data samples
enriched in jets misidentified as photons are selected, either by inverting the photon isolation
variable ($\mathcal{I}>0.07$), or by requiring at least one track in a cone of ${\cal R}<0.05$ around the
photon, while keeping the remaining photon selection criteria unchanged. In both cases the agreement
with the MC prediction for $\varepsilon_{\rm jet}$ is found to be within 10\%, which is taken as the
systematic uncertainty.
The total number of DPP events
is found to be $N_{\gamma\gamma} =7307 \pm 312({\rm stat.})$,  corresponding to an average DPP purity
of $\approx 67\%$. Following this procedure, the number
of DPP events is estimated in each bin of the four kinematic variables considered (\mgg, \ptgg, \dphigg, and \cosgg).
The largest kinematic dependence of the DPP purity is in terms of \mgg,
with a variation between $\approx 60\%$ at $M_{\gamma\gamma}\approx 40$~GeV and
close to 100\% for $M_{\gamma\gamma}> 200$~GeV. As a function
of the other kinematic variables, the DPP purity varies in the ($60 - 70$)\% range.
The relative systematic uncertainty on the purity results from the systematic uncertainties
on $\varepsilon_{\gamma}$ and $\varepsilon_{\rm jet}$, and typically varies within ($11 - 15$)\%.
As a cross-check, the DPP purity was also estimated via a fit to the two-dimensional distribution
in data of $O_{\rm NN, \gamma_1}$ versus $O_{\rm NN, \gamma_2}$ using templates constructed
from photons and jets in MC. The result was found to be in good agreement with that from
the $4\times 4$ matrix method.

The estimated number of DPP events per bin is corrected for the DPP event selection efficiency and acceptance.
The selection efficiency is calculated using DPP events generated
with {\sc pythia} and processed through a {\sc geant}-based simulation
of the D0 detector. In order to accurately model the effects of multiple $p\bar{p}$ interactions
and detector noise, data events from random $p\bar{p}$ crossings with a similar instantaneous
luminosity spectrum as considered in the data analysis are overlaid on the MC events. These MC events
are then processed using the same reconstruction code as for the data. Small differences between
data and MC in the per-photon selection efficiencies are corrected for with suitable scale factors
derived using control samples of electrons from $Z$ boson decays, as well as photons from
the radiative $Z$ boson decays.
The overall DPP selection efficiency after applying all selection criteria is estimated
as a function of the variable of interest. In the case of \ptgg, \dphigg, and \cosgg,
it is about 64\% with a (2 - 3)\% variation
across the bins, while for \mgg, the efficiency grows from about 60\% at $30<M_{\gamma\gamma}<50$~GeV
to 69\% at $M_{\gamma\gamma}>200$~GeV.
The total relative systematic uncertainty on the DPP selection efficiency is 4.3\%,
dominated by the track-match veto and photon $O_{\rm NN}$ selections.
The acceptance is calculated using DPP events generated with {\sc resbos} and
is driven by the selections in $\eta_{\rm det}$ ($|\eta_{\rm det}|<0.9$, applied to avoid
edge effects in the central calorimeter region used for the measurement) and $\phi_{\rm det}$
(to avoid periodic calorimeter module boundaries~\cite{d0det}
that bias the EM cluster energy and position measurements),
PV misidentification, photon energy scale, and bin-to-bin migration effects due
to the finite energy and angular resolution of the EM calorimeter. The overall DPP acceptance
varies within ($45 - 64$)\% with a relative systematic uncertainty of ($4 - 7$)\%.

The differential cross sections \Dmgg, \Dptgg, \Ddphigg, and \Dcosgg~are
obtained from the number of data events corrected for the background contribution,
divided by the trigger, vertex and diphoton selection efficiencies, acceptance,
integrated luminosity, and the bin width for each kinematic variable.
The measured differential cross sections, compared to the theoretical predictions from {\sc resbos},
are presented in Table \ref{tab:sigmas_1d}.
The average value for each variable in a bin was estimated using {\sc resbos}.
The statistical uncertainty $\delta_{\rm stat}$ corresponds to the statistical
precision on $N_{\gamma\gamma}$ estimated in the $4\times 4$ matrix method,
which can be sizable when values of $\epsilon_\gamma$ and $\epsilon_{\rm jet}$
are numerically close.

\begin{figure*}[htbp]
 \centering
\hspace*{-2mm} \includegraphics[scale=0.27]{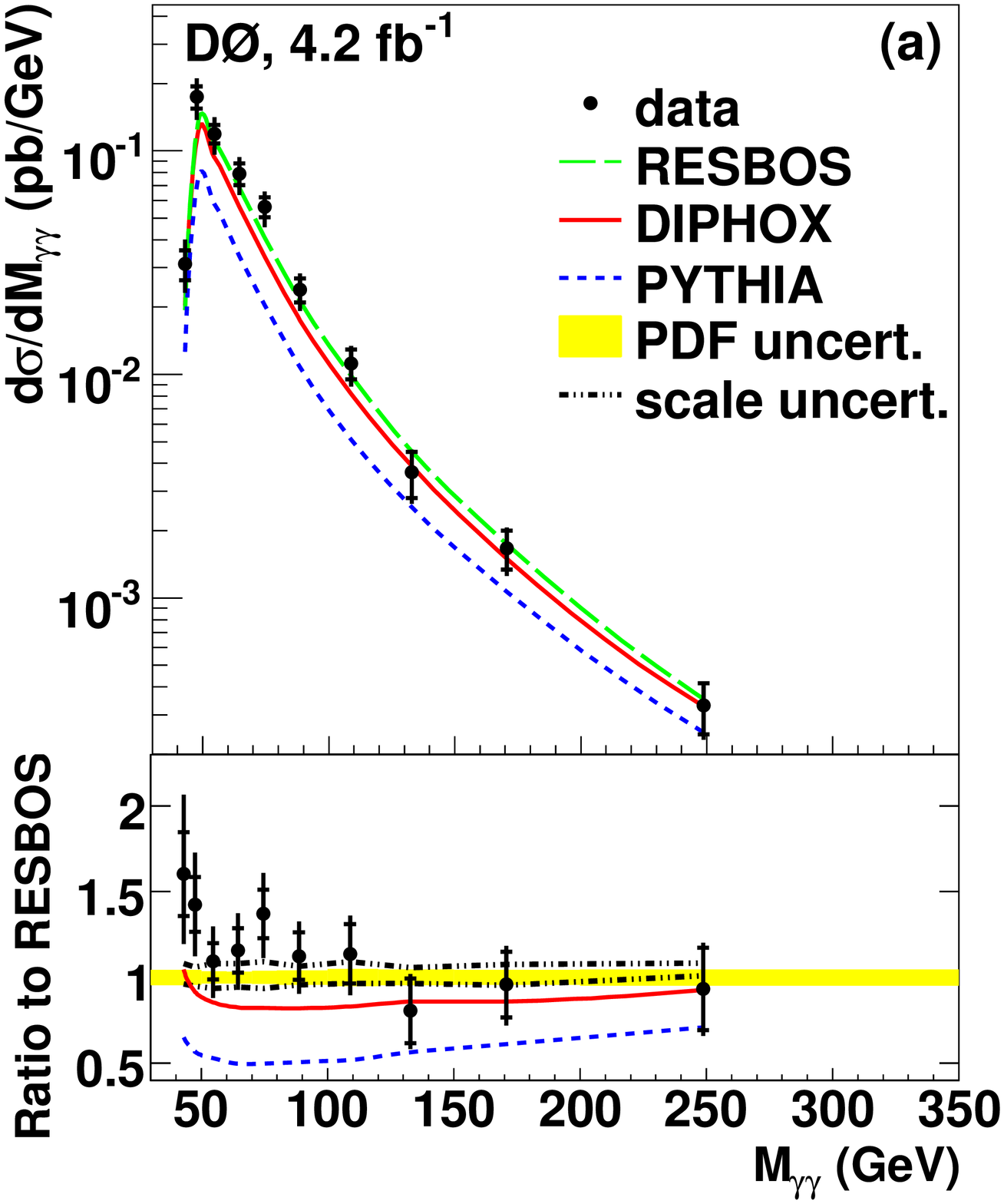}
\hspace*{5mm}  \includegraphics[scale=0.27]{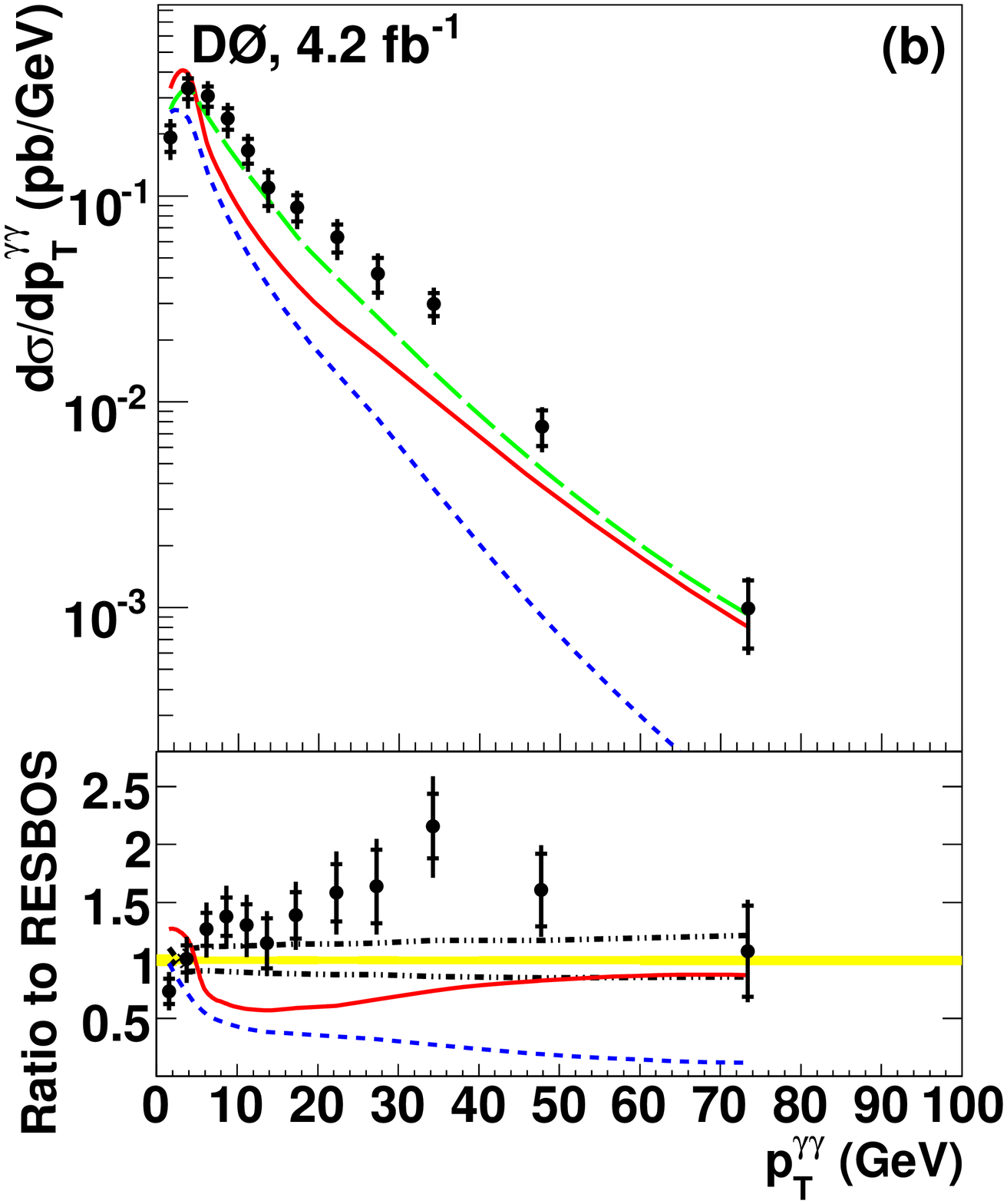}
~\\[0mm]
\hspace*{-2mm}  \includegraphics[scale=0.27]{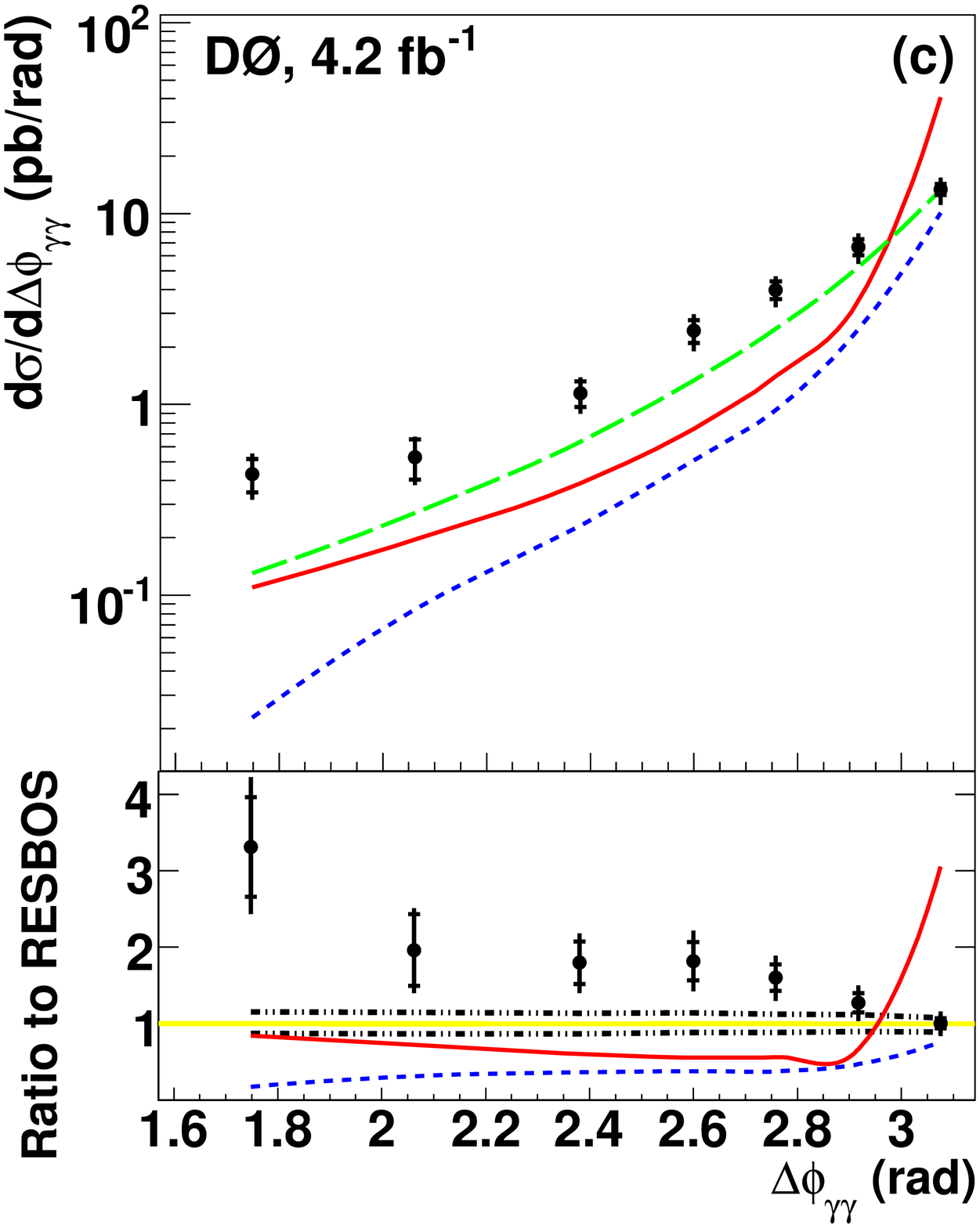}
\hspace*{5mm}  \includegraphics[scale=0.27]{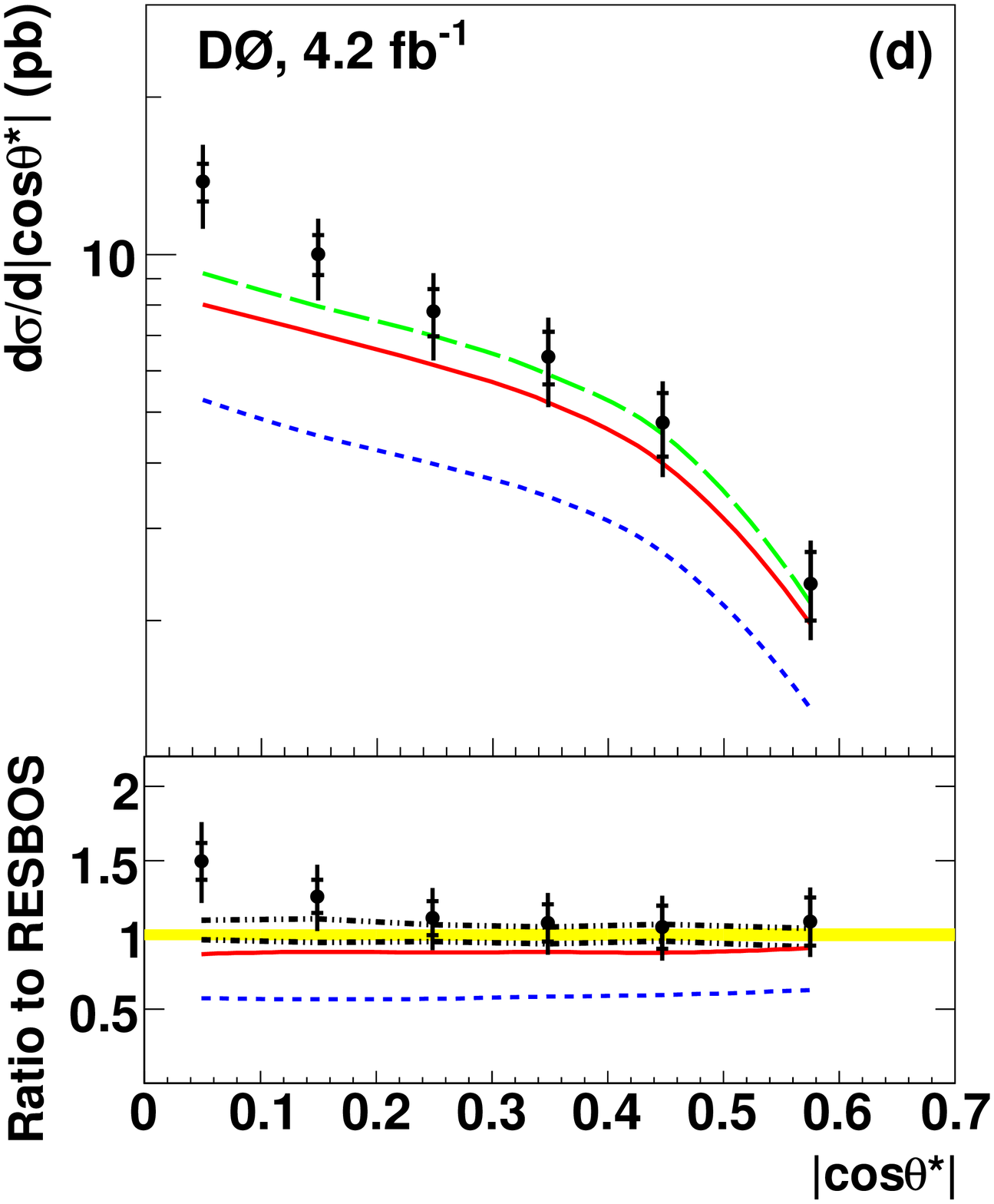}
~\\[-5mm]
\caption{The measured differential diphoton production cross sections as functions of
(a) \mgg, (b) \ptgg, (c) \dphigg, and (d) \cosgg. The data are compared to the theoretical predictions from
{\sc resbos}, {\sc diphox}, and {\sc pythia}.
The predictions from {\sc resbos}, and {\sc diphox} use the CTEQ6.6M PDF set~\cite{CTEQ}
and renormalization, factorization, and fragmentation scales $\mu_{R} = \mu_{F} = \mu_f = $ \mgg, while
{\sc pythia} uses the Tune A settings.
Theoretical predictions are obtained using the following selections: two photons with $p_T>21(20)$ GeV, $|\eta|<0.9$,
$30<$\mgg$<350$ GeV, \mgg$>p_{T}^{\gamma\gamma}$, ${\Delta\cal{R}}>0.4$, $\Delta \phi_{\gamma \gamma}>0.5\pi$, and
$E^{\text {iso}}_{T}<2.5$ GeV.
The ratio  of differential cross sections between data and {\sc resbos} are displayed as black points
with uncertainties in the bottom plots.
The inner line for the uncertainties in data points shows the statistical uncertainty, while the outer line shows
the total (statistical and systematic added in quadrature) uncertainty
after removing the 7.4\% normalization uncertainty.
The solid (dashed) line shows the ratio of the predictions from {\sc diphox} ({\sc pythia}) to those from {\sc resbos}.
In the bottom plots, the scale uncertainties are shown by dash-dotted lines and the PDF uncertainties by shaded regions.}
\label{fig:xsec_ratio}
\end{figure*}

Figure~\ref{fig:xsec_ratio} shows a comparison of the measured differential cross sections to the
theoretical predictions from {\sc resbos}, {\sc diphox}, and {\sc pythia}.
Systematic uncertainties in the measured cross sections have large ($>\!90\%$) bin-to-bin
correlations.  There is a common $7.4\%$ normalization uncertainty,
resulting from the photon selection criteria (4.3\%)
and luminosity measurement (6.1\%), that is not shown on the data points.
The predictions from {\sc resbos} and {\sc diphox} are computed using the CTEQ6.6M PDF set~\cite{CTEQ},
the DSS set of fragmentation functions \cite{DSS_ff}, and
setting renormalization $\mu_{R}$, factorization $\mu_F$, and fragmentation $\mu_f$ scales as
$\mu_{R}=\mu_{F}=\mu_f=M_{\gamma\gamma}$.
The uncertainty due to the scale choice is estimated by simultaneous variation by a factor of two of
all scales relative to the default choice and found to be about 10\% for \mgg~and \cosgg~and up to ($15 - 20$)\%
for high \ptgg~and low \dphigg.
The PDF uncertainty is estimated using {\sc diphox} and
the 44 eigenvectors provided with the CTEQ6.6M PDF set~\cite{CTEQ} and
found to be within ($3 - 6$)\% for all four cross sections.
The predictions from {\sc pythia} are computed with ``Tune A'' \cite{pythia-ref}, which uses the CTEQ5L PDF set.
All theoretical predictions are obtained using diphoton event selection criteria
equivalent to those applied in the experimental analysis. In particular, the photon isolation
is required to be $E_{T}^{\rm iso} =E_{T}^{\rm tot}(0.4)-E_{T}^\gamma < 2.5$ GeV,
where $E_{T}^{\rm tot}(0.4)$ is the total transverse energy
within a cone of radius ${\cal R}=0.4$ centered on the photon, and  $E_{T}^\gamma$ is the photon transverse energy.
For {\sc resbos} and {\sc diphox}, $E_{T}^{\rm tot}$ is computed at the parton level, whereas in the
case of {\sc pythia}, it is computed at the particle level.
This requirement suppresses the contributions from photons produced in the fragmentation processes
and leads to a more consistent comparison with the experimental result.
Studies performed using {\sc diphox} indicate that the contribution to the overall cross section
from one- and two-fragmentation processes
does not exceed 16\% and significantly drops at large \mgg, \ptgg ~and small \dphigg to (1--3)\%.
In order to allow a direct comparison to the data, the NLO QCD cross sections obtained with {\sc resbos} and {\sc diphox}
are further corrected for contributions from multiple parton interactions and hadronization, both of which
affect the efficiency of the isolation requirement.
These corrections are estimated using DPP events simulated in {\sc pythia} using Tunes A and S0 \cite{pythia-ref}.
The corrections vary within ($4.0 - 5.5$)\% as a function of the measured kinematic variables
and are consistent for both tunes within 0.5\%.

The results obtained show that none of the theoretical predictions considered is able to describe the data well in all
kinematic regions of the four variables. {\sc resbos} shows the best agreement with data,  although
systematic discrepancies are observed at low \mgg, high \ptgg, and low \dphigg.  However, the agreement between
{\sc resbos} and data is fair at intermediate \mgg~($50 - 80$~GeV), and good at high \mgg~($>80$~GeV).
The large discrepancy between {\sc resbos} and {\sc diphox}
in some regions of the phase space is due
to absence of all-order soft-gluon resummation and accounting
$gg\to\gamma\gamma$ contribution just at LO in {\sc diphox}.

\begin{table}[htbp]
\begin{center}
\small
\caption{The measured differential cross sections in bins of \mgg, \ptgg, \dphigg, and \cosgg. The columns
$\delta_{\text {stat}}$ and $\delta_{\text {syst}}$ represent the statistical and systematic uncertainties, respectively.
Also shown are the predictions from {\sc resbos}.}
\label{tab:sigmas_1d}
\begin{tabular}{l@{\hspace{-0.6mm}}c@{\hspace{-0.4mm}}c@{\hspace{-1.0mm}}c@{\hspace{1.5mm}}c@{\hspace{1.5mm}}c} \hline\hline
$\quad M_{\gamma\gamma}$ &$\langle M_{\gamma\gamma}\rangle$ & \multicolumn{4}{c}{$d\sigma/dM_{\gamma\gamma}$ (pb/GeV) }\\\cline{3-6}
 \hspace{0.6mm} (GeV) & (GeV)& Data & $\delta_{\text {stat}}$ (\%) & $\delta_{\text {syst}}$ (\%) & {\sc resbos}  \\\hline
  30 -- 45 & 43.0 & 3.11$\times10^{-2}$ & 15 & +26/$-29$& 1.94$\times10^{-2}$ \\
  45 -- 50 &  47.6 &   1.74$\times10^{-1}$ &  11 &+19/$-19$ & 1.22$\times10^{-1}$     \\
  50 -- 60 &  54.7 &   1.19$\times10^{-1}$ &  10 &+18/$-17$ & 1.09$\times10^{-1}$     \\
  60 -- 70 &  64.6 &   7.89$\times10^{-2}$ &  11 &+18/$-16$ & 6.82$\times10^{-2}$     \\
  70 -- 80 &  74.6 &   5.61$\times10^{-2}$ &  10 &+17/$-15$ & 4.09$\times10^{-2}$     \\
  80 -- 100 & 88.6 &  2.39$\times10^{-2}$ &  12 &+16/$-15$& 2.13$\times10^{-2}$     \\
 100 -- 120 & 108.9 &  1.12$\times10^{-2}$ &  15 &+16/$-14$& 0.98$\times10^{-2}$     \\
 120 -- 150 & 132.9 &  3.65$\times10^{-3}$ &  23 &+16/$-14$& 4.52$\times10^{-3}$     \\
 150 -- 200 & 170.7 &  1.67$\times10^{-3}$ & 20 &+16/$-14$ & 1.74$\times10^{-3}$     \\
 200 -- 350 & 248.8 & 3.30$\times10^{-4}$ & 26 &+16/$-14$ & 3.53$\times10^{-4}$     \\\hline\\\hline
 \hspace{3.8mm}$p_{T}^{\gamma\gamma}$ & $\langle p_{T}^{\gamma\gamma}\rangle$ & \multicolumn{4}{c}{$d\sigma/dp_{T}^{\gamma\gamma}$ (pb/GeV)} \\\cline{3-6}
  \hspace{0.6mm} (GeV) &(GeV)& Data& $\delta_{\text {stat}}$ (\%) & $\delta_{\text {syst}}$ (\%)&{\sc resbos}  \\\hline
  0.0 -- 2.5 &   1.5 &  1.92$\times10^{-1}$ &  15 &+18/$-19$ &  2.63$\times10^{-1}$    \\
  2.5 -- 5.0 &  3.7 &   3.34$\times10^{-1}$ &  11 &+19/$-17$ &  3.30$\times10^{-1}$    \\
  5.0 -- 7.5 &  6.2 &   3.06$\times10^{-1}$ &  11 &+17/$-16$ &  2.41$\times10^{-1}$    \\
  7.5 -- 10.0 & 8.7 &   2.38$\times10^{-1}$ &  12 &+18/$-17$ &  1.73$\times10^{-1}$    \\
  10.0 -- 12.5 &11.2 &  1.66$\times10^{-1}$ &  14 &+18/$-16$ &  1.28$\times10^{-1}$    \\
  12.5 -- 15.0 & 13.7 & 1.10$\times10^{-1}$ &  19 &+18/$-17$ &  9.57$\times10^{-2}$    \\
  15.0 --  20.0 &  17.3 & 8.80$\times10^{-2}$ & 15 &+18/$-17$ & 6.34$\times10^{-2}$    \\
  20.0 --  25.0 &  22.3 & 6.30$\times10^{-2}$ & 16 &+18/$-18$ & 3.98$\times10^{-2}$    \\
  25.0 --  30.0 &  27.3 & 4.20$\times10^{-2}$ & 19 &+18/$-18$ & 2.57$\times10^{-2}$    \\
  30.0 --  40.0 &  34.3 & 2.99$\times10^{-2}$ & 13 &+18/$-17$ & 1.39$\times10^{-2}$    \\
  40.0 --  60.0 &  47.8 & 7.58$\times10^{-3}$ & 20 &+17/$-16$ & 4.72$\times10^{-3}$    \\
  60.0 -- 100 &  73.4 & 9.92$\times10^{-4}$ & 36 &+19/$-21$ & 9.20$\times10^{-4}$    \\\hline\\\hline
\hspace{4.4mm}$\Delta\phi_{\gamma\gamma}$ & $\langle \Delta\phi_{\gamma\gamma}\rangle$ & \multicolumn{4}{c}{$d\sigma/d\Delta\phi_{\gamma\gamma}$ (pb/rad)} \\\cline{3-6}
  \hspace{3.6mm} (rad) & (rad)& Data  & $\delta_{\text {stat}}$ (\%) & $\delta_{\text {syst}}$ (\%)& {\sc resbos}  \\\hline
1.57 -- 1.88 & 1.75 &   4.32$\times10^{-1}$ &   20 &+19/$-21$ &   1.31$\times10^{-1}$    \\
1.88 -- 2.20 & 2.06 &   5.30$\times10^{-1}$ &   24 &+18/$-16$ &   2.70$\times10^{-1}$    \\
2.20 -- 2.51 & 2.38 &   1.15 &   16 &+18/$-16$ &   6.38$\times10^{-1}$    \\
2.51 -- 2.67 & 2.60 &   2.43 &   14 &+19/$-19$ &   1.34    \\
2.67 -- 2.83 & 2.76 &   3.99 &   11 &+17/$-16$ &   2.49    \\
2.83 -- 2.98 & 2.92 &   6.70 &    10 &+18/$-16$ &   5.25    \\
2.98 -- 3.14 & 3.08 &  1.34$\times10^{1}$ &    7 &+17/$-16$ &  1.33$\times10^{1}$    \\\hline \\\hline
\hspace{2.4mm}\cosgg & $\langle$\cosgg$\rangle$ & \multicolumn{4}{c}{$d\sigma/d$\cosgg (pb)} \\\cline{3-6}
    & & Data& $\delta_{\text {stat}}$ (\%) & $\delta_{\text {syst}}$ (\%)  & {\sc resbos}  \\\hline
 0.0 -- 0.1 &  0.05 &   13.8 &    8 &+18/$-17$ &   9.22    \\
 0.1 -- 0.2 &  0.15 &   10.0 &    9 &+17/$-16$ &   7.96    \\
 0.2 -- 0.3 &  0.25 &   7.78 &   10 &+18/$-16$ &   6.99    \\
 0.3 -- 0.4 &  0.35 &   6.38 &   12 &+17/$-16$ &   5.90    \\
 0.4 -- 0.5 &  0.45 &   4.77 &   14 &+17/$-16$ &   4.54    \\
 0.5 -- 0.7 &  0.57 &   2.35 &   15 &+17/$-16$ &   2.16    \\\hline \hline

\end{tabular}
\end{center}
\end{table}

\begin{figure*}[htbp]
 \centering
\hspace*{-5mm} \includegraphics[scale=0.27]{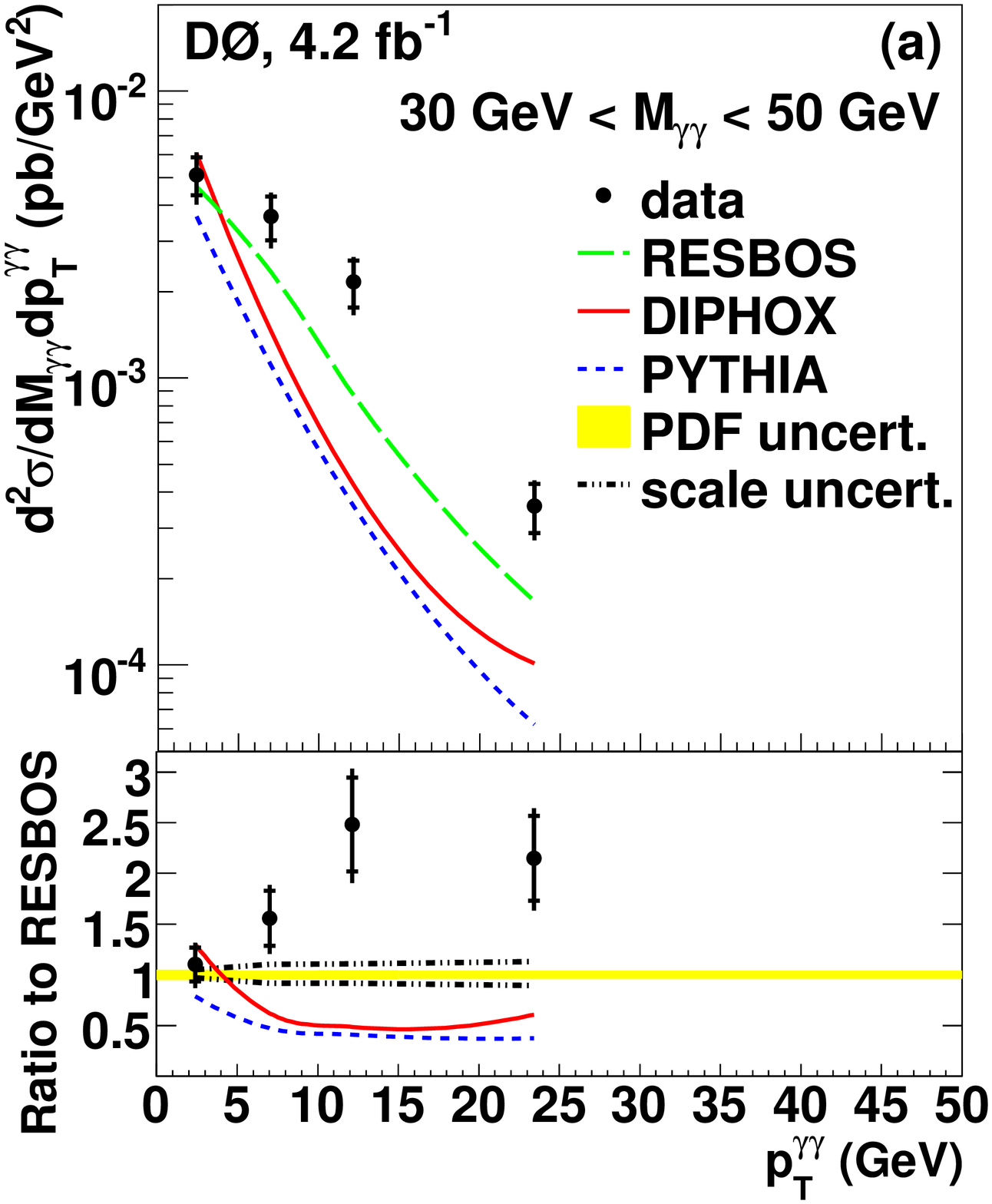}
\hspace*{2mm}  \includegraphics[scale=0.27]{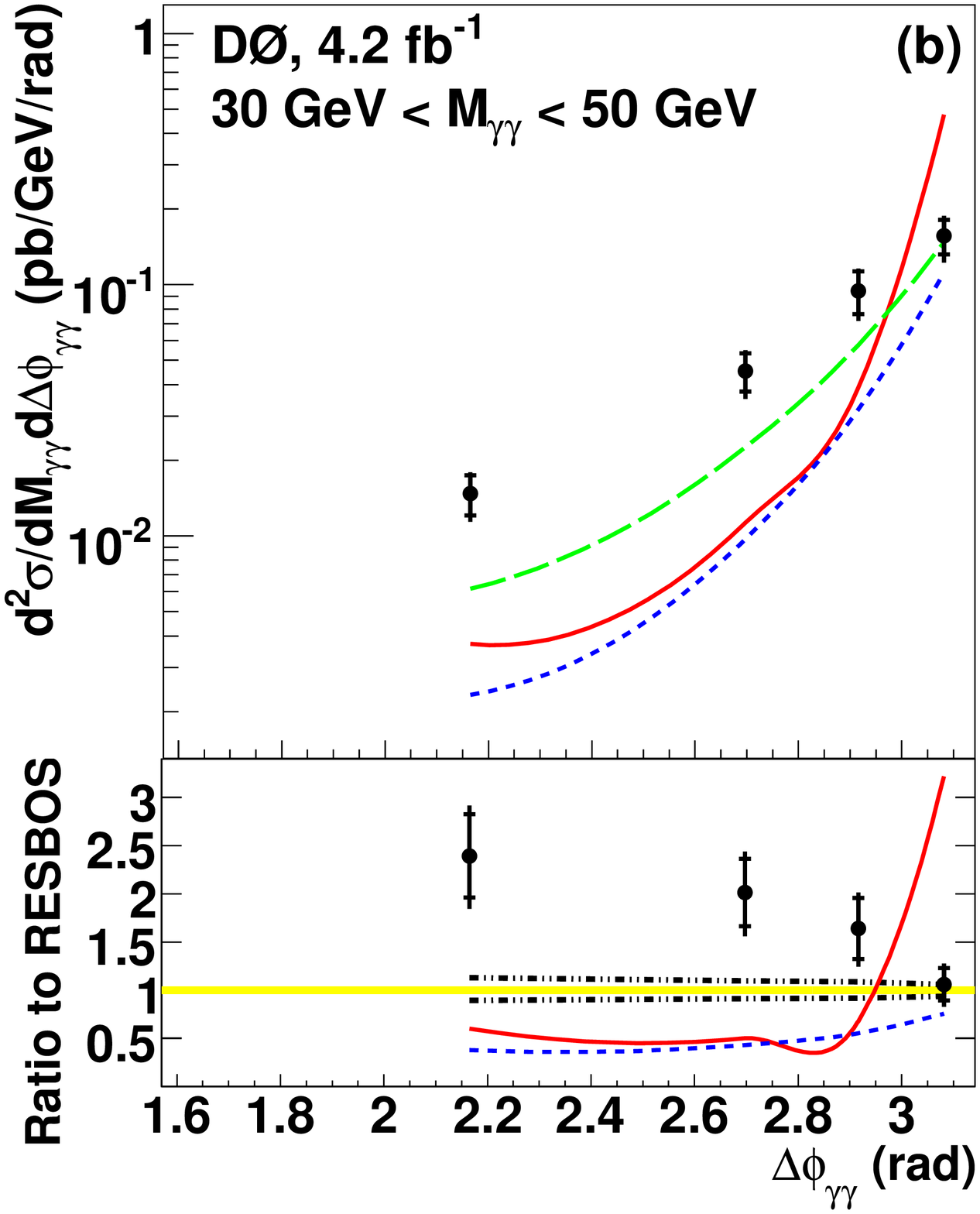}
\hspace*{2mm}  \includegraphics[scale=0.27]{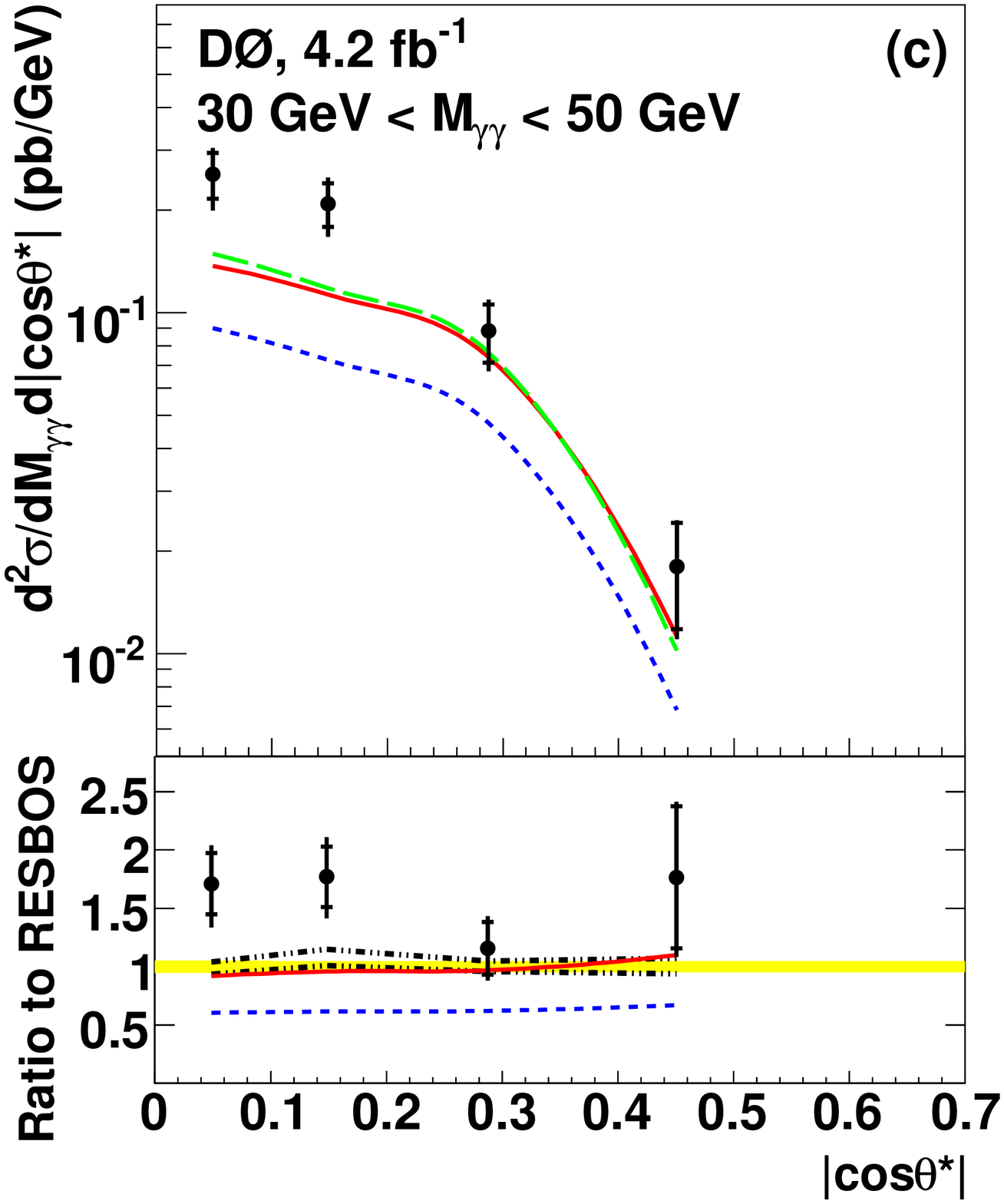}
~\\[-7mm]
\caption{The measured double differential diphoton production cross sections as functions of
(a) \ptgg, (b) \dphigg, and (c) \cosgg ~for $30<$~\mgg~$<50$ GeV.
The notations for points, lines and shaded regions are the same as in Figure~\ref{fig:xsec_ratio}. }
\label{fig:xsec_ratio_3050}
\end{figure*}

\begin{figure*}[htbp]
 \centering
\hspace*{-5mm} \includegraphics[scale=0.27]{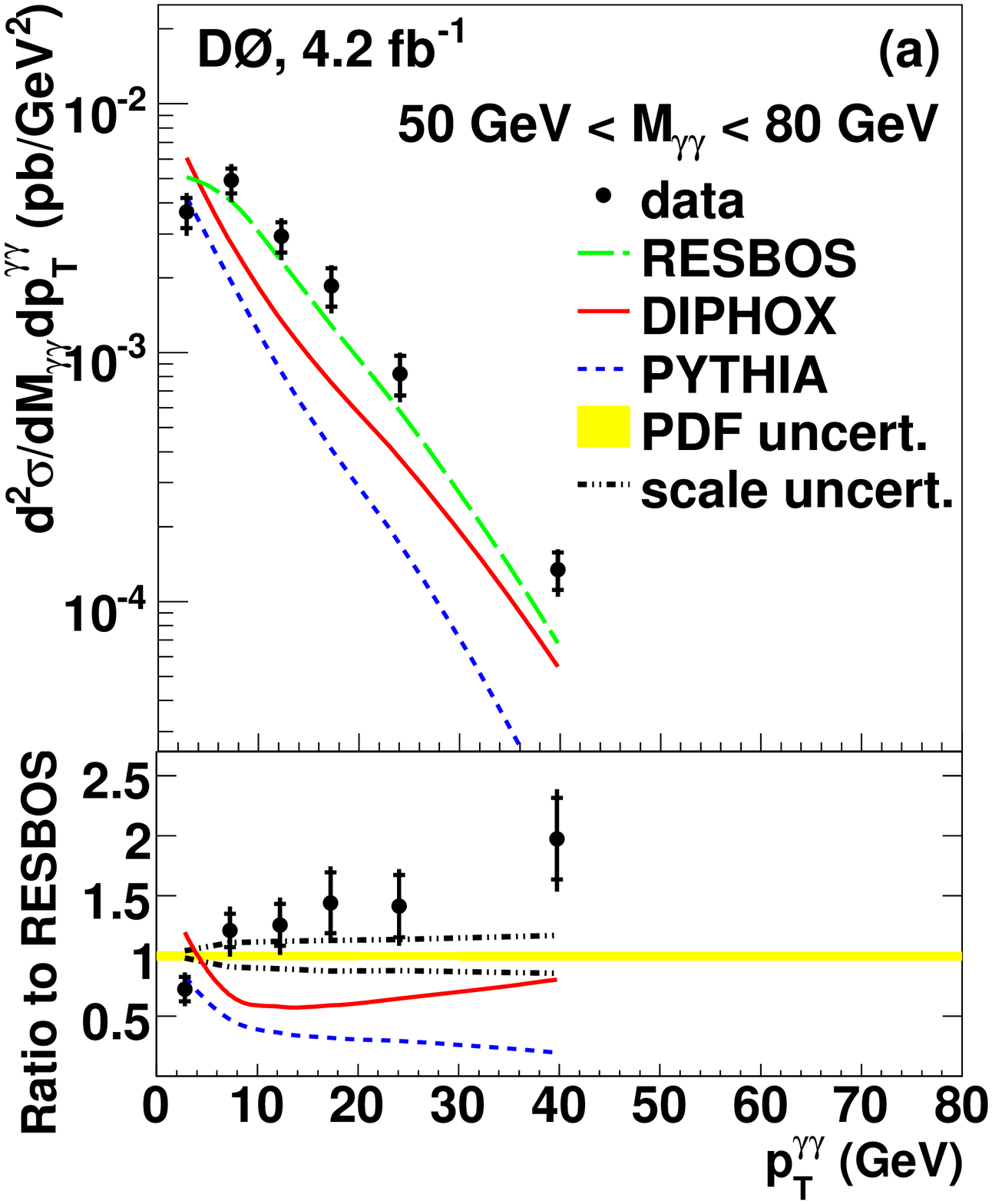}
\hspace*{2mm}  \includegraphics[scale=0.27]{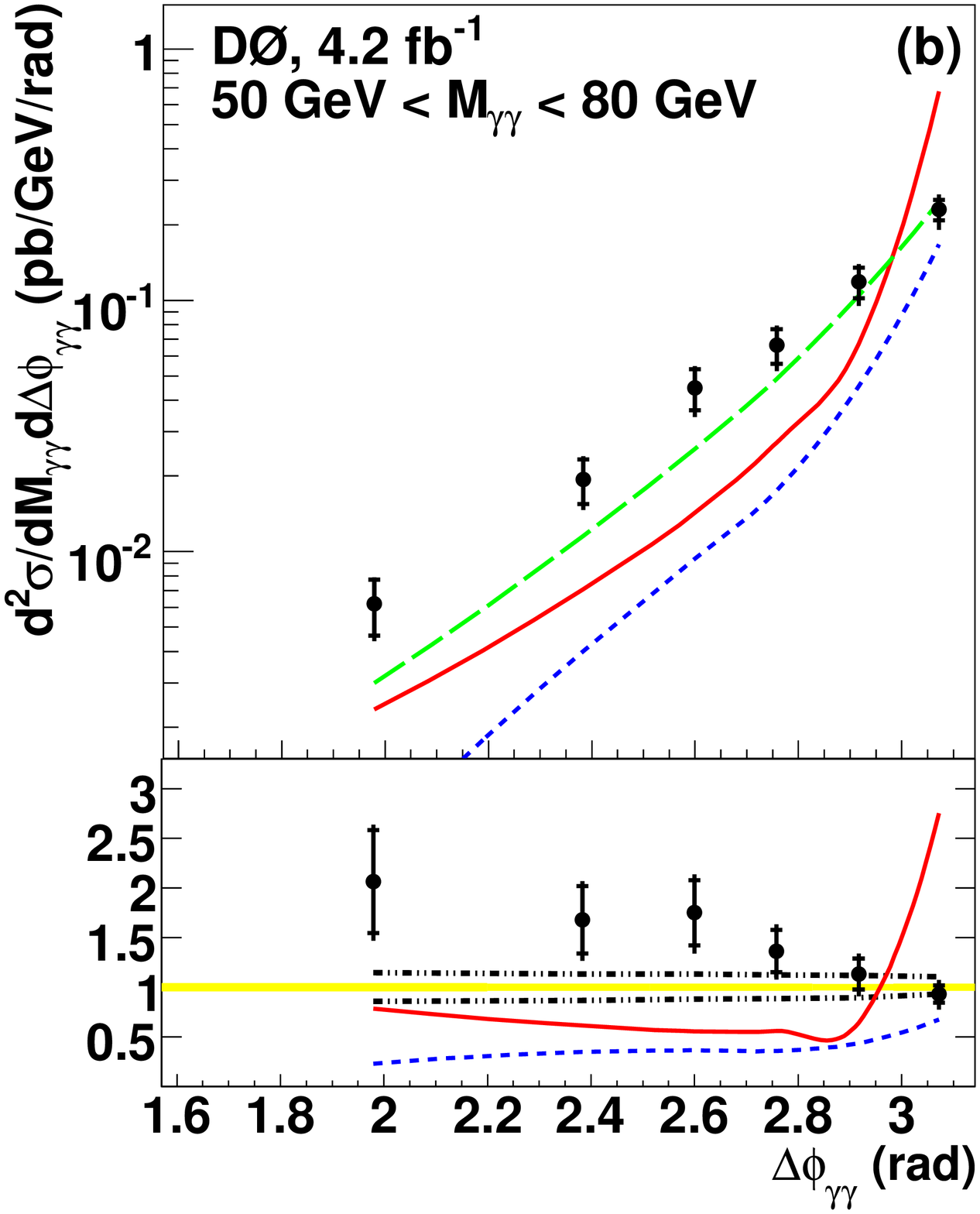}
\hspace*{2mm}  \includegraphics[scale=0.27]{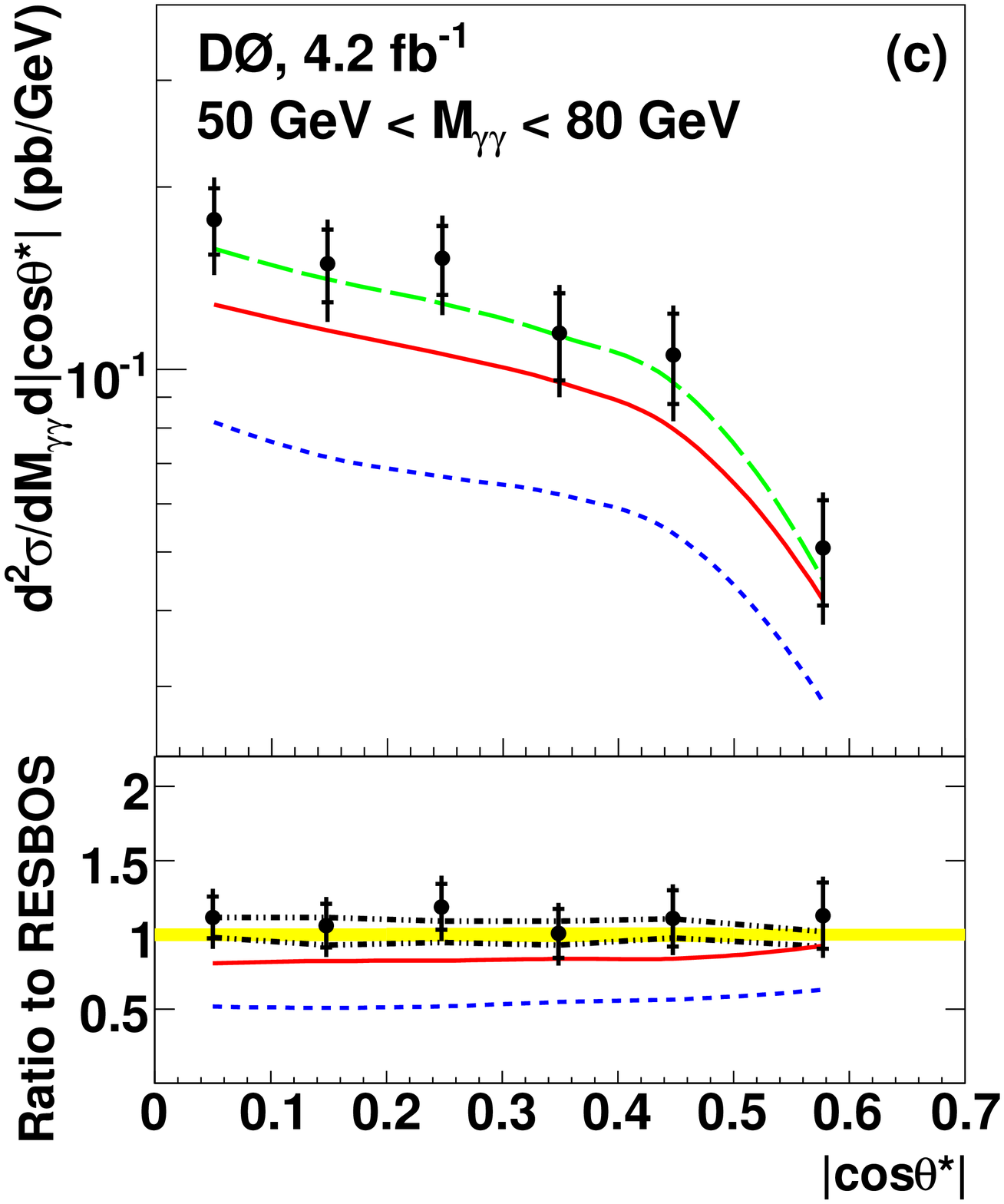}
~\\[-7mm]
\caption{The measured double differential diphoton production cross sections as functions of
(a) \ptgg, (b) \dphigg, and (c) \cosgg ~for $50<$~\mgg~$<80$ GeV.
The notations for points,  lines and shaded regions are the same as in Figure~\ref{fig:xsec_ratio}. }
\label{fig:xsec_ratio_5080}
\end{figure*}

\begin{figure*}[htbp]
 \centering
\hspace*{-5mm} \includegraphics[scale=0.27]{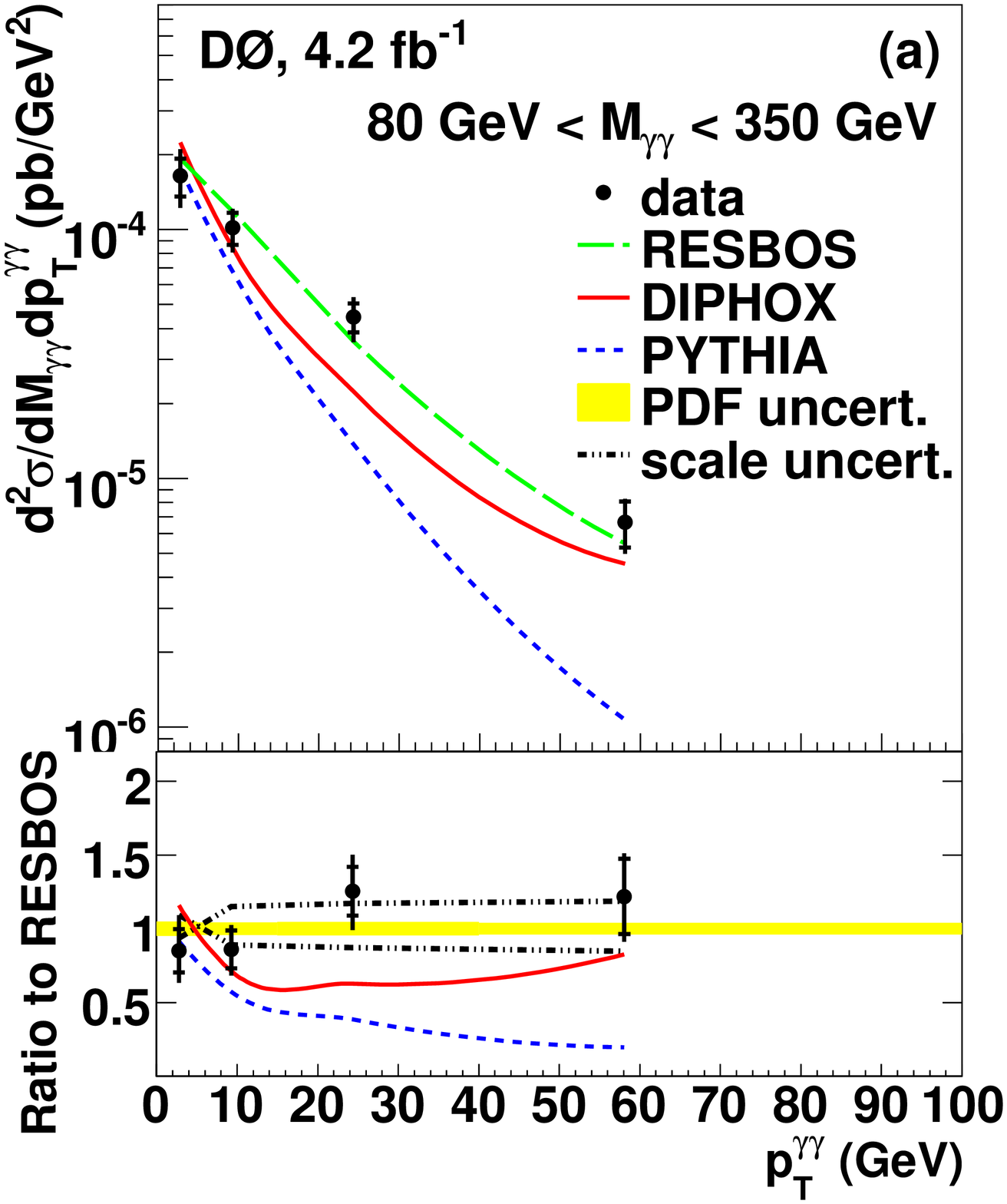}
\hspace*{2mm}  \includegraphics[scale=0.27]{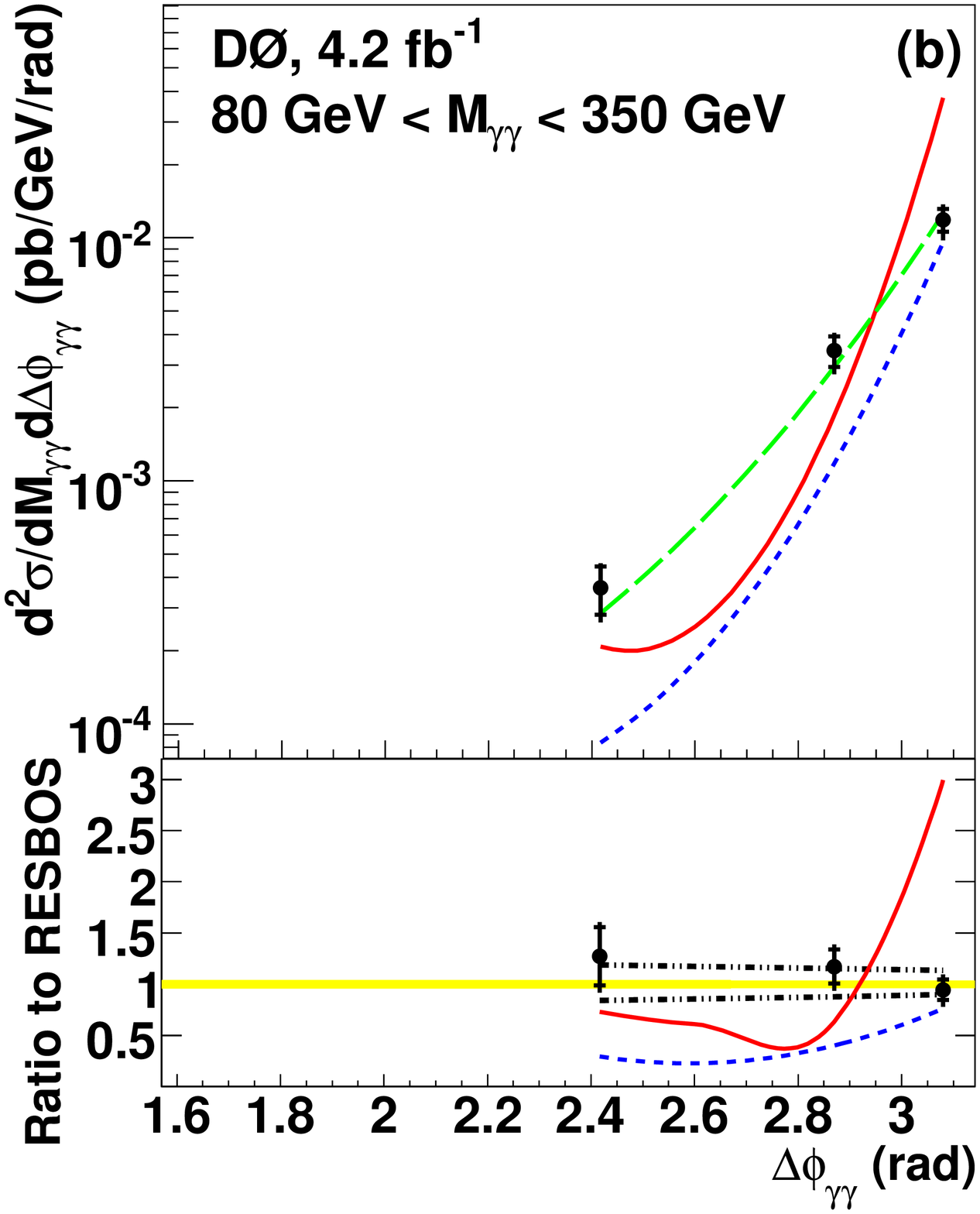}
\hspace*{2mm}  \includegraphics[scale=0.27]{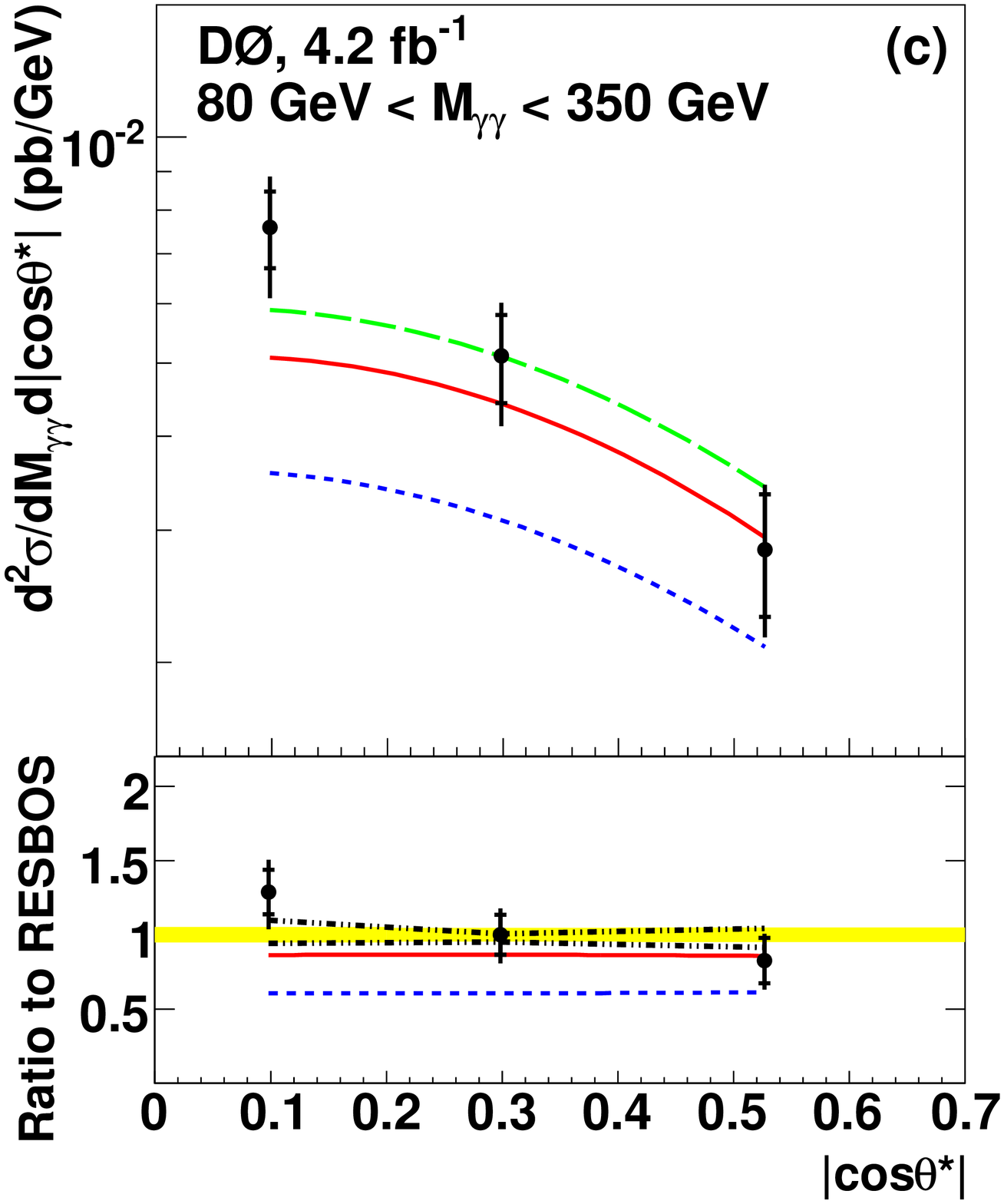}
~\\[-7mm]
\caption{The measured double differential diphoton production cross sections as functions of
(a) \ptgg, (b) \dphigg, and (c) \cosgg ~for $80<$~\mgg~$<350$ GeV.
The notations for  points, lines and shaded regions are the same as in Figure~\ref{fig:xsec_ratio}. }
\label{fig:xsec_ratio_80350}
\end{figure*}

Further insight on the dependence of the \ptgg, \dphigg, and \cosgg ~kinematic distributions on the
mass scale can be gained through the measurement of double differential cross sections.
For this purpose, the differential cross sections as functions of \ptgg, \dphigg, and \cosgg ~are measured in three
\mgg ~bins: $30 - 50$ GeV, $50 - 80$ GeV and $80 - 350$ GeV.
The results are presented in Tables \ref{tab:sigma_2D_m3050} -- \ref{tab:sigma_2D_m80350},
corresponding to each of the three \mgg ~intervals.
Each table is split into three sub-tables, showing results separately for \DTptgg, \DTdphigg, and \DTcosgg.
The measured cross sections for the \ptgg, \dphigg, and \cosgg ~variables in the three mass bins
are shown in Figures~\ref{fig:xsec_ratio_3050} $-$ \ref{fig:xsec_ratio_80350}
and compared to the theoretical predictions.
These results confirm that the largest discrepancies
between data and {\sc resbos} for each of the kinematic variables originate from the lowest \mgg~region
(\mgg~$<50$~GeV). As shown in Figure~\ref{fig:ggF}, this is the region where the contribution from
$gg\to\gamma\gamma$ is expected to be largest. The discrepancies between data and {\sc resbos} are
reduced in the intermediate \mgg~region ($50 - 80$ GeV), and a quite satisfactory description
of all kinematic variables is achieved for the \mgg$>80$~GeV region, the
relevant region for the Higgs boson and new phenomena searches.
However, it should be pointed out that at the Tevatron, DPP production
at high masses is strongly dominated by $q\bar{q}$ annihilation, in contrast
with the LHC, where the contribution from $gg$ and $qg$ initiated process will be significant.
It remains to be seen whether the addition of
NNLO corrections to {\sc resbos}, as done in \cite{HWW}, will improve the description of
the high \ptgg ~(low \dphigg) spectrum at low \mgg.

In summary, we have presented measurements of single and double differential cross sections for
DPP production in $p\bar{p}$ collisions at $\sqrt{s}=1.96$~TeV. This analysis
uses 4.2 fb$^{-1}$ of D0 data, representing a twenty-fold increase in statistics relative to the
last published Tevatron results \cite{CDF}. The measured cross sections are compared
to predictions from {\sc resbos}, {\sc diphox} and {\sc pythia},
showing the necessity of including higher order corrections
beyond NLO as well as the resummation to all orders of soft and collinear initial state gluons.
These results allow the tuning of the theoretical predictions for this process, which
is of great relevance for improving the sensitivity of searches for the Higgs boson and other
new phenomena at the Tevatron and the LHC.

\begin{table}[htbp]
\begin{center}
\small
\caption{The measured double differential cross sections in bins of \ptgg, \dphigg, and \cosgg, in the
region $30 < M_{\gamma\gamma} < 50$~GeV. The columns $\delta_{\text {stat}}$ and $\delta_{\text {syst}}$ represent the
statistical and systematic uncertainties, respectively. Also shown are the predictions from {\sc resbos}.}
\label{tab:sigma_2D_m3050}
\vskip 1mm
\begin{tabular}{l@{\hspace{0mm}}c@{\hspace{-0.7mm}}c@{\hspace{-1.0mm}}c@{\hspace{1.5mm}}c@{\hspace{1.5mm}}c} \hline\hline
\hspace{2.8mm} $p_{T}^{\gamma\gamma}$ & $\langle p_{T}^{\gamma\gamma}\rangle$ & \multicolumn{4}{c}{$d^2\sigma/dM_{\gamma\gamma}dp_{T}^{\gamma\gamma}$ (pb/GeV$^{2}$)} \\\cline{3-6}
  \hspace{0.6mm} (GeV) & (GeV)& Data & $\delta_{\text {stat}}$ (\%) & $\delta_{\text {syst}}$ (\%) & {\sc resbos}  \\\hline

   0.0 --   5.0 &   2.4 &    5.11$\times10^{-3}$ &   15 & +17/$-14$ &    4.64$\times10^{-3}$    \\
   5.0 --  10.0 &   7.0 &    3.65$\times10^{-3}$ &   18 & +16/$-14$ &    2.35$\times10^{-3}$    \\
  10.0 --  15.0 &  12.2 &    2.17$\times10^{-3}$ &   19 & +16/$-14$ &    8.72$\times10^{-4}$    \\
  15.0 --  50.0 &  23.4 &    3.58$\times10^{-4}$ &   19 & +16/$-14$ &    1.67$\times10^{-4}$    \\\hline\\\hline

 \hspace{3.8mm} $\Delta\phi_{\gamma\gamma}$ & $\langle \Delta\phi_{\gamma\gamma}\rangle$ & \multicolumn{4}{c}{$d^2\sigma/dM_{\gamma\gamma}d\Delta\phi_{\gamma\gamma}$ (pb/GeV/rad)} \\\cline{3-6
}
  \hspace{3.6mm} (rad) & (rad)& Data & $\delta_{\text {stat}}$ (\%) & $\delta_{\text {syst}}$ (\%) & {\sc resbos}  \\\hline

1.57 -- 2.51 &  2.16 &   1.48$\times10^{-2}$ &  18 & +16/$-14$ &   6.16$\times10^{-3}$    \\
2.51 -- 2.83 &  2.70 &   4.54$\times10^{-2}$ &   17 & +16/$-14$ &   2.25$\times10^{-2}$    \\
2.83 -- 2.98 &  2.92 &   9.45$\times10^{-2}$ &   19 & +16/$-14$ &   5.76$\times10^{-2}$    \\
2.98 -- 3.14 &  3.08 &   1.57$\times10^{-1}$ &   16 & +16/$-14$ &   1.48$\times10^{-1}$    \\\hline\\\hline

 \hspace{1.6mm} $|\cos\theta^{*}|$ & $\langle |\cos\theta^{*}|\rangle$ & \multicolumn{4}{c}{$d^2\sigma/dM_{\gamma\gamma}d|\cos\theta^{*}|$ (pb/GeV)} \\\cline
{3-6}
    & & Data & $\delta_{\text {stat}}$ (\%) & $\delta_{\text {syst}}$ (\%) & {\sc resbos}  \\\hline

 0.0 -- 0.1 &  0.05 &   2.55$\times10^{-1}$ &   15 & +17/$-14$ &   1.49$\times10^{-1}$    \\
 0.1 -- 0.2 &  0.15 &   2.09$\times10^{-1}$ &   15 & +16/$-15$ &   1.18$\times10^{-1}$    \\
 0.2 -- 0.4 &  0.28 &   8.84$\times10^{-2}$ &   19 & +16/$-16$ &   7.64$\times10^{-2}$    \\
 0.4 -- 0.7 &  0.44 &   1.80$\times10^{-2}$ &   35 & +19/$-15$ &   1.02$\times10^{-2}$    \\\hline\hline

\end{tabular}
\end{center}
\end{table}

\begin{table}[htbp]
\begin{center}
\small
\caption{The measured double differential cross sections in bins of \ptgg, \dphigg, and \cosgg, in the
region $50 < M_{\gamma\gamma} < 80$~GeV. The notations are  the same as in Table \ref{tab:sigma_2D_m3050}.
}
\label{tab:sigma_2D_m5080}
\vskip 1mm
\begin{tabular}{l@{\hspace{0mm}}c@{\hspace{-0.7mm}}c@{\hspace{-1.0mm}}c@{\hspace{1.5mm}}c@{\hspace{1.5mm}}c} \hline\hline
\hspace{2.8mm} $p_{T}^{\gamma\gamma}$ & $\langle p_{T}^{\gamma\gamma}\rangle$ & \multicolumn{4}{c}{$d^2\sigma/dM_{\gamma\gamma}dp_{T}^{\gamma\gamma}$ (pb/GeV$^{2}$)} \\\cline{3-6
}
  \hspace{0.6mm} (GeV) & (GeV)& Data &$\delta_{\text {stat}}$ (\%) & $\delta_{\text {syst}}$ (\%) & {\sc resbos}  \\\hline

   0.0 --   5.0 &   2.8 &    3.68$\times10^{-3}$ &   14 &+16/$-15$ &    5.07$\times10^{-3}$    \\
   5.0 --  10.0 &   7.3 &    4.92$\times10^{-3}$ &   12 &+16/$-14$ &    4.06$\times10^{-3}$   \\
  10.0 --  15.0 &  12.3 &    2.93$\times10^{-3}$ &   14 &+16/$-14$ &    2.33$\times10^{-3}$   \\
  15.0 --  20.0 &  17.3 &    1.86$\times10^{-3}$ &   18 &+16/$-14$ &    1.29$\times10^{-3}$   \\
  20.0 --  30.0 &  24.1 &    8.22$\times10^{-4}$ &   18 &+16/$-14$ &    5.81$\times10^{-4}$   \\
  30.0 --  80.0 &  39.8 &    1.34$\times10^{-4}$ &   17 &+16/$-14$ &    6.81$\times10^{-5}$   \\\hline\\\hline

 \hspace{3.8mm} $\Delta\phi_{\gamma\gamma}$ & $\langle \Delta\phi_{\gamma\gamma}\rangle$ & \multicolumn{4}{c}{$d^2\sigma/dM_{\gamma\gamma}d\Delta\phi_{\gamma\gamma}$ (pb/GeV/rad)} \\\cline{3-6
}
  \hspace{3.6mm} (rad) & (rad)& Data & $\delta_{\text {stat}}$ (\%) & $\delta_{\text {syst}}$ (\%)  & {\sc resbos}  \\\hline
1.57 -- 2.20 &  1.98 &   6.19$\times10^{-3}$ &   25 &+16/$-14$ &   2.99$\times10^{-3}$    \\
2.20 -- 2.51 &  2.38 &   1.94$\times10^{-2}$ &   20 &+16/$-14$ &   1.16$\times10^{-2}$    \\
2.51 -- 2.67 &  2.60 &   4.49$\times10^{-2}$ &   19 &+16/$-14$ &   2.56$\times10^{-2}$    \\
2.67 -- 2.83 &  2.76 &   6.64$\times10^{-2}$ &   16 &+16/$-14$ &   4.87$\times10^{-2}$    \\
2.83 -- 2.98 &  2.92 &   1.18$\times10^{-1}$ &   14&+16/$-14$ &    1.04$\times10^{-1}$   \\
2.98 -- 3.14 &  3.07 &   2.30$\times10^{-1}$ &    10 &+16/$-14$ &  2.47$\times10^{-1}$    \\\hline\\\hline

 \hspace{1.6mm} $|\cos\theta^{*}|$ & $\langle |\cos\theta^{*}|\rangle$ & \multicolumn{4}{c}{$d^2\sigma/dM_{\gamma\gamma}d|\cos\theta^{*}|$ (pb/GeV)} \\\cline
{3-6}
    & & Data & $\delta_{\text {stat}}$ (\%) & $\delta_{\text {syst}}$ (\%) & {\sc resbos}  \\\hline

 0.0 -- 0.1 &  0.05 &   1.77$\times10^{-1}$ &   13 &+16/$-14$ &   1.58$\times10^{-1}$    \\
 0.1 -- 0.2 &  0.15 &   1.50$\times10^{-1}$ &   14 &+16/$-14$ &   1.41$\times10^{-1}$    \\
 0.2 -- 0.3 &  0.25 &   1.53$\times10^{-1}$ &   13 &+16/$-14$ &   1.29$\times10^{-1}$    \\
 0.3 -- 0.4 &  0.35 &   1.15$\times10^{-1}$ &   16 &+16/$-14$ &   1.14$\times10^{-1}$    \\
 0.4 -- 0.5 &  0.45 &   1.06$\times10^{-1}$ &   17 &+16/$-14$ &   9.52$\times10^{-2}$   \\
 0.5 -- 0.7 &  0.58 &   5.08$\times10^{-2}$ &   20 &+17/$-14$ &   4.50$\times10^{-2}$    \\\hline\hline

\end{tabular}
\end{center}
\end{table}

\begin{table}[htbp]
\begin{center}
\small
\caption{The measured double differential cross sections in bins of \ptgg, \dphigg, and \cosgg, in the
region $80 < M_{\gamma\gamma} <350$~GeV. The notations are  the same as in Table \ref{tab:sigma_2D_m3050}.
}
\label{tab:sigma_2D_m80350}
\vskip 1mm
\begin{tabular}{l@{\hspace{0mm}}c@{\hspace{-0.7mm}}c@{\hspace{-1.0mm}}c@{\hspace{1.5mm}}c@{\hspace{1.5mm}}c} \hline\hline
\hspace{2.8mm} $p_{T}^{\gamma\gamma}$ & $\langle p_{T}^{\gamma\gamma}\rangle$ & \multicolumn{4}{c}{$d^2\sigma/dM_{\gamma\gamma}dp_{T}^{\gamma\gamma}$ (pb/GeV$^{2}$)} \\\cline{3-6
}
  \hspace{0.6mm} (GeV) & (GeV)& Data & $\delta_{\text {stat}}$ (\%) & $\delta_{\text {syst}}$ (\%) & {\sc resbos}  \\\hline
  0.0 -- 5.0 &   2.8 &    1.64$\times10^{-4}$ &   17 &+20/$-24$ &    1.93$\times10^{-4}$    \\
  5.0 --15.0 &   9.3 &    1.02$\times10^{-4}$ &   15 &+16/$-14$ &    1.18$\times10^{-4}$    \\
 15.0 --40.0 &  24.3 &    4.46$\times10^{-5}$ &   13 &+18/$-16$ &    3.56$\times10^{-5}$    \\
 40.0 --100 &  58.1 &   6.67$\times10^{-6}$ &  21 &+16/$-14$ &   5.48$\times10^{-6}$    \\\hline\\\hline

 \hspace{3.8mm} $\Delta\phi_{\gamma\gamma}$ & $\langle \Delta\phi_{\gamma\gamma}\rangle$ & \multicolumn{4}{c}{$d^2\sigma/dM_{\gamma\gamma}d\Delta\phi_{\gamma\gamma}$ (pb/GeV/rad)} \\\cline{3-6
}
  \hspace{3.6mm} (rad) & (rad)& Data & $\delta_{\text {stat}}$ (\%) & $\delta_{\text {syst}}$ (\%) & {\sc resbos}  \\\hline
1.57 -- 2.67 &  2.42 &   3.63$\times10^{-4}$ &   22 &+18/$-16$ &   2.85$\times10^{-4}$    \\
2.67 -- 2.98 &  2.87 &   3.44$\times10^{-3}$ &   14 &+16/$-14$ &   2.94$\times10^{-3}$    \\
2.98 -- 3.14 &  3.08 &   1.19$\times10^{-2}$ &   11 &+16/$-14$ &   1.26$\times10^{-2}$   \\\hline\\\hline

 \hspace{1.6mm} $|\cos\theta^{*}|$ & $\langle |\cos\theta^{*}|\rangle$ & \multicolumn{4}{c}{$d^2\sigma/dM_{\gamma\gamma}d|\cos\theta^{*}|$ (pb/GeV)} \\\cline
{3-6}
    & & Data & $\delta_{\text {stat}}$ (\%) & $\delta_{\text {syst}}$ (\%) & {\sc resbos}  \\\hline
 0.0 -- 0.2 &  0.10 &   7.58$\times10^{-3}$ &   12& +17/$-14$ &   5.89$\times10^{-3}$    \\
 0.2 -- 0.4 &  0.30 &   5.11$\times10^{-3}$ &   13& +16/$-14$ &   5.11$\times10^{-3}$    \\
 0.4 -- 0.7 &  0.53 &   2.82$\times10^{-3}$ &   19& +16/$-14$ &   3.42$\times10^{-3}$    \\\hline\hline

\end{tabular}
\end{center}
\end{table}

%We thank C.~Balazs, C.-P.~Yuan and J.P.~Guillet for useful discussions and their
%assistance with the theoretical predictions.
We thank C.~Balazs, C.-P.~Yuan and J.P.~Guillet for their
assistance with the theoretical predictions.
We also thank F.~Siegert and S.~Schumann for useful discussions.
% acknowledgement_paragraph_r2.tex                          2/4/10
%
We thank the staffs at Fermilab and collaborating institutions, 
and acknowledge support from the 
DOE and NSF (USA);
CEA and CNRS/IN2P3 (France);
FASI, Rosatom and RFBR (Russia);
CNPq, FAPERJ, FAPESP and FUNDUNESP (Brazil);
DAE and DST (India);
Colciencias (Colombia);
CONACyT (Mexico);
KRF and KOSEF (Korea);
CONICET and UBACyT (Argentina);
FOM (The Netherlands);
STFC and the Royal Society (United Kingdom);
MSMT and GACR (Czech Republic);
CRC Program and NSERC (Canada);
BMBF and DFG (Germany);
SFI (Ireland);
The Swedish Research Council (Sweden);
and
CAS and CNSF (China).

\bibliography{reference}

\begin{thebibliography}{33}
\bibitem{d0_coordinate}
The polar angle $\theta$ and the azimuthal angle $\phi$ are defined 
with respect to the positive $z$ axis, which is along the proton beam direction. 
Pseudorapidity is defined as $\eta=-\ln[\tan(\theta/2)]$.
Also, $\eta_{\text {det}}$ and $\phi_{\text {det}}$ are the pseudorapidity and the azimuthal angle 
measured with respect to the center of the detector.
\end{thebibliography}
\bibliographystyle{apsrev}
\end{document}